\documentclass[%
 reprint,
superscriptaddress,
 amsmath,amssymb,
prx,
]{revtex4-2}

\usepackage[T1]{fontenc}
\usepackage{graphicx}
\usepackage{comment}

\usepackage{dcolumn}
\usepackage{bm}

\usepackage{lineno}
\usepackage{float}
\usepackage{booktabs}
\usepackage{siunitx}

\usepackage{chemformula}
\usepackage{tabularx} 
\usepackage{textcomp}
\DeclareSIUnit\year{yr}

\begin{document}

\preprint{APS/123-QED}
\pagenumbering{Roman}

\title{Cold Hypersonic Gas Jets for Precision Laser Spectroscopy of Radioactive Molecules}

\author{A. Claessens}
\altaffiliation[Present address: ]{Facility for Rare Isotope Beams, Michigan state university, East Lansing 48824 MI, USA}
\email{Contact author: claessen@frib.msu.edu}
\affiliation{KU Leuven, Instituut voor Kern- en Stralingsfysica, 3001 Leuven, Belgium}

\date{\today}

\author{R. Ferrer}
 \email{Contact author: rafael.ferrer@kuleuven.be}
\affiliation{KU Leuven, Instituut voor Kern- en Stralingsfysica, 3001 Leuven, Belgium}

\author{Harshithbabu}
\affiliation{KU Leuven, Instituut voor Kern- en Stralingsfysica, 3001 Leuven, Belgium}

\author{L. M. Arndt}
\affiliation{Johannes Gutenberg University Mainz, 55099 Mainz,
	Germany}
\affiliation{Helmholtz-Institut Mainz, 55099 Mainz,
	Germany}

\author{Ch. E. D\"ullmann}
\affiliation{Johannes Gutenberg University Mainz, 55099 Mainz,
	Germany}
\affiliation{Helmholtz-Institut Mainz, 55099 Mainz,
	Germany}
\affiliation{GSI Helmholtzzentrum für Schwerionenforschung GmbH, 64291 Darmstadt, Germany}

\author{K. Gaul}
\affiliation{Helmholtz-Institut Mainz, 55099 Mainz,
	Germany}
\affiliation{GSI Helmholtzzentrum für Schwerionenforschung GmbH, 64291 Darmstadt, Germany}
\affiliation{Johannes Gutenberg University Mainz, 55099 Mainz,
	Germany}

\author{F. Ivandikov}
\affiliation{KU Leuven, Instituut voor Kern- en Stralingsfysica, 3001 Leuven, Belgium}
\author{G. Neyens}
\affiliation{KU Leuven, Instituut voor Kern- en Stralingsfysica, 3001 Leuven, Belgium}

\author{D. Renisch}
\affiliation{Johannes Gutenberg University Mainz, 55099 Mainz,
	Germany}
\affiliation{Helmholtz-Institut Mainz, 55099 Mainz,
	Germany}
\author{J. Stricker}
\affiliation{Johannes Gutenberg University Mainz, 55099 Mainz,
	Germany}
\affiliation{Helmholtz-Institut Mainz, 55099 Mainz,
	Germany}	
\author{W. Saelens}
\affiliation{KU Leuven, Instituut voor Kern- en Stralingsfysica, 3001 Leuven, Belgium}
\author{P. Van den Bergh}
\affiliation{KU Leuven, Instituut voor Kern- en Stralingsfysica, 3001 Leuven, Belgium}
\author{P. Van Duppen}
\affiliation{KU Leuven, Instituut voor Kern- en Stralingsfysica, 3001 Leuven, Belgium}

\begin{abstract}
The sensitivity to possible beyond standard model physics via a study of the nuclear Schiff moment (NSM) has since recently been driving laser spectroscopy of short-lived radioactive molecules. These experiments have in part been hindered by inefficient cooling of the molecular degrees of freedom, resulting in complex, difficult to interpret molecular spectra. Hypersonic (de Laval) nozzles have been shown to provide ideal environments to study cold short-lived actinides by means of in-gas-jet laser ionization spectroscopy. We present here the first application of this technique on molecules using $^{229,230,232}$ThO laser ablation targets. Translational and rotational temperatures were characterized via laser spectroscopy and found to be $\SI{12.5\pm 2.5}{\kelvin}$ and $\SI{7.2 \pm 0.6}{\kelvin}$, respectively, along the nozzle-collimated $\SI{60}{\mm}$-long, $\SI{5}{\mm}$-diameter gas jet. An upper bound of the vibrational temperature was estimated as no transitions from $v''=1$ were observed. Molecular King-plot analysis shows a large field shift of $-\SI{4.0\pm 1.3}{\cm^{-1} /fm^2}$, indicating enhanced sensitivity to the differential nuclear charge radii. Relativistic FSCCSD calculations reproduce the measured molecular field-shift factor, providing a benchmark between state-of-the-art molecular and atomic calculations.

\end{abstract}

\maketitle


\section{Introduction}
Radioactive molecule studies offer unique opportunities in the fields of fundamental physics, astrophysics, nuclear structure and chemistry. The quantum control properties of molecular states have allowed for some of the strongest constraints on possible beyond standard (BSM) model theories via electric dipole moment (EDM) measurements \cite{Engel2013}. Combined with exotic short-lived nuclei, radioactive molecules are predicted to be extremely sensitive probes to charge conjugation and parity violating (CPV) moments in the low-energy regime of the strong interaction \cite{Yu2021,Arrowsmith-kron2024}.  Besides BSM searches, radioactive molecules provide prospects for astro- and nuclear physics. Short to medium term dynamics in stellar evolution could be tracked using the microwave emission of cold radioactive tracer molecules \cite{Kaminski2018,Arrowsmith-kron2024}. On the other hand, molecular enhancements could unlock greater sensitivity to nuclear structure effects, e.g., the Bohr-Weisskopf effect \cite{Wilkins2025}. Progress in radioactive ion beam production and laser spectroscopy sensitivity now enables the study of these molecules (e.g. at ISOLDE CERN \cite{GarciaRuiz2020}).
\\
\\
In the nucleus, nucleon EDMs and CPV internuclear forces will mix odd-parity nuclear states resulting in a permanent nuclear EDM. In a neutral atom or molecule, this moment is completely shielded by the electrons  \cite{Schiff1963} and the CP violation effect is instead revealed through the nuclear Schiff moment (NSM) in systems with nuclear spin $>0$. The internal electric field in the nuclear volume created by the core-penetrating valence electrons interacts with the NSM, contributing to the atomic EDM or the molecular EDM \cite{Safronova2018, Chupp2019}. While the NSM is the leading order nuclear contribution, a multitude of different and complementary EDM measurements are needed to constrain all the possible sources of CPV \cite{Chupp2015,Degenkolb2026, Gaul2024}. The current best limits on hadronic CPV sources are set by the atomic EDM measurements on $^{199}$Hg \cite{Graner2016} and $^{129}$Xe \cite{Sachdeva2019,Allmendinger2019}. The use of a radioactive molecule can greatly increase the sensitivity to the NSM. First of all, the NSM is strongly enhanced ($10^2$ - $10^6$) in nuclei with octupole deformation (e.g. specific isotopes of Fr, Ra, Th and Pa) compared to Hg \cite{Arrowsmith-kron2024, Gaffney2013}. Furthermore, easily polarized by an external field, a molecule provides an experimental platform containing a strong electric field that couples to the EDM \cite{Safronova2018}. Molecules such as $^{232}$ThO \cite{Andreev2018} and $^{180}$HfF$^+$ \cite{Roussy2023} already provide the most stringent limits on the electron EDM, probing energy scales outside of the Large Hadron Collider's reach. 
 \\
 \\
 Many radioactive molecules such as $^{227}$AcF \cite{ AthanasakisKaklamanakis2025}, $^{225}$RaOCH$_3^+$ \cite{Yu2021,Gaul2024a}, $^{229}$PaF$^{3+}$ \cite{Zulch2022}, $^{227,229}$ThF$^+$, and $^{227,229}$ThO \cite{Barker2012,Skripnikov2015,Denis2015,Liang1995,Hammond2002, Flambaum2019}, have been proposed for NSM searches. Two of these candidates are isotopologues of thorium monoxide: $^{227,229}$ThO where the $^{227,229}$Th isotopes have nuclear half-lives of $19$~days and $7900$~years, respectively. Their short half-lives and limited availability either constrain the sample size to the microgram range and below at off-line facilities or require on-line production (i.e., directly coupled to an accelerator beamline) at heavy-ion accelerator facilities, which in both cases demands a sensitive and precise spectroscopy technique. The In-Gas Laser Ionization and Spectroscopy (IGLIS) technique is optimized for the study of actinides produced in minute quantities by fusion-evaporation reactions \cite{Ferrer2017,Granados2017}. Its gas-jet cooling addresses a major obstacle encountered in recent pioneering experiments on short-lived molecules conducted at ISOLDE-CERN. In these experiments (cf. \cite{GarciaRuiz2020, AthanasakisKaklamanakis2025}), short-lived radioactive molecules of RaF and AcF from proton-induced reactions and in-target fluorination \cite{Au2023} were studied by Collinear Resonance Ionization Spectroscopy using the CRIS setup. Such studies indicate that the molecules have high internal temperatures that originate from a combination of RIB production and ion manipulation, which compromises the sensitivity and increases spectral congestion. We report here the extension of the IGLIS technique to molecular species produced by a laser-ablation source and demonstrate its application through the study of the isotopologue chain $^{229,230,232}$ThO using in-gas-jet laser spectroscopy \cite{Ferrer2017}.  A convergent-divergent (de Laval) hypersonic nozzle creates a well-collimated and cold gas jet without the need for skimmers or electrostatic lenses. The efficient cooling of the molecular degrees of freedom is characterized using  $^{232}$ThO molecules (Sec.~\ref{sec:translationalTemp}). Thorium molecules are produced by laser ablation of a metallic ($^{232}$Th) target or, for the less abundant isotopes, a salt-based target (see Sec.~\ref{sec:targets}), inside an argon-filled gas cell. Rotational constants of $^{230,232}$ThO are extracted from the analysis of the data as well as the isotopologue shifts of $^{229,230,232}$ThO (Sec. \ref{sec:results}). Molecular spectroscopic parameters are also obtained from relativistic ab initio electronic-structure calculations (Sec. \ref{sec:theory}) and discussed along with the experimental results in Sec. \ref{sec:outlook}.

 \section{In-gas-jet spectroscopy of thorium molecules}\label{sec:Temperatures}
 \begin{figure*}
 	\includegraphics{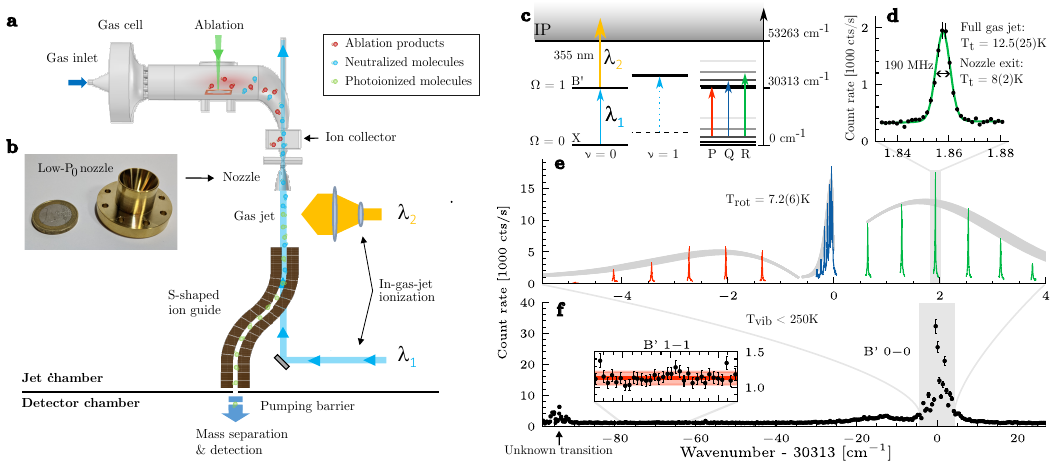}
 	\caption{Schematic layout of the IGLIS setup ($\mathbf{a}$) at KU Leuven. Ablation of a ThO target in a high-purity argon atmosphere ($P_0 = \SI{80}{\milli \bar}$) results in an ensemble of hot molecular ions and neutral molecules. The continuous flow carries and thermalizes the products on their way to an ion collector ($\SI{-50}{\volt}$) located near the exit. Neutral molecules are extracted via a contoured de Laval nozzle ($\mathbf{b}$) where the hypersonic flow regime allows for efficient high-resolution resonance ionization spectroscopy. The flow is characterized by a high Mach number and uniform density over a $\SI{60}{\mm}$ distance, indicating a high degree of collimation at the best flow match conditions, i.e., $P_0 = \SI{80}{\milli \bar}$ and $P_\text{bkg} = \SI{5.9e-3}{\milli \bar}$. Several different operating conditions for the nozzle were also measured in ref.~\cite{thesisClaessens}. ThO is laser-ionized by a two-step ionization scheme as shown in $\mathbf{c}$, after which the photo-ions are captured by an S-shaped radio-frequency quadrupole (RFQ) ion guide and brought to a high vacuum region for mass separation and detection. The resulting spectra show the cooling capacity of the contoured nozzle: the velocity spread (translational), i.e, $T_t$ is cooled to $\SI{8 \pm 2}{\kelvin}$ at the nozzle exit as measured in the B'-X R(2) transition $\mathbf{d}$. A temperature of $\SI{12.5 \pm 2.5}{\kelvin}$ was measured over the full gas jet (see text). The full B'-X rovibrational transition is measured and shown in $\mathbf{e}$. The intensity band in grey indicates the obtained rotational temperature of $\SI{7.2 \pm 0.6}{\kelvin}$. No higher vibrational states ($\nu''> 0$) are occupied with a measurable population ($\mathbf{f}$) as observed in the zoomed $\nu''=1 \rightarrow \nu'=1$ region. The average fluctuation of the background and $1\sigma$ band is shown in red. Taking two standard deviations, an upper bound of $\SI{250}{\kelvin}$ for the vibrational degree is found. At around  $\SI{30219}{\per \cm}$ in $\mathbf{f}$, an unknown transition is visible, revealed by the reduced vibrational spectral congestion.}\label{fig:iglissetup_mainfigure}  
 \end{figure*}
The molecule ThO has a well-known structure \cite{Edvinsson1987,Edvinsson1988,Goncharov2005,Long2014,Schmitz2019} and its natural abundant isotopologue $^{232}$ThO was shown to be produced easily using ablation in a gas cell filled with argon \cite{Claessens2025}. ThO isotopologues thus provide an excellent opportunity to characterize the in-gas-jet method regarding both temperature (spectral resolution and congestion) and efficiency, with the aim of pushing towards shorter-lived isotopes and eventually on-line implementation. In this section, we present the IGLIS setup at KU Leuven adapted for off-line molecular studies and characterize the cooling properties of the hypersonic nozzle on $^{232}$ThO molecules. A schematic of the experimental setup is depicted in Fig.~\ref{fig:iglissetup_mainfigure}. 
\\
\\ 
Either metallic or salt-based ablation targets are placed in a gas cell with continuous argon flow at a pressure of $\SI{80}{\milli \bar}$ to thermalize and transport the ablation products towards the gas cell exit. The ion collectors remove left-over (molecular) ions before extraction out of the cell via a contoured hypersonic nozzle. Two-step resonance ionization spectroscopy (RIS) is then performed, after which the photo-ions are captured by an S-shaped radiofrequency quadrupole (RFQ) and guided through several differential pumping stages to a high-voltage region for mass separation and ion detection. Two different nozzles for hypersonic flow (i.e. with a Mach number $M>5$) have been designed, manufactured and characterized using atoms \cite{zadvornaya2018characterization,Ferrer2021}. Both of the nozzles create well-collimated, uniform gas jets over a distance of at least $\SI{60}{\mm}$, a diameter of $\SI{5}{\mm}$ and an average M of around $8.5$. Their application is dependent on the gas cell pressure (i.e., stagnation pressure $P_0$). A low-$P_0$ nozzle \cite{Lantis2024} with a design $P_0$ of $\SI{80}{\milli \bar}$ is used for gas cells with recoil sources \cite{Claessens2025} or fast-extraction gas cells making use of electric fields \cite{Raeder2020,Lantis2024} where $P_0$ is limited to $\lesssim\SI{100}{\milli \bar}$. On the other hand, for only-flow gas cells pressures of $200-\SI{500}{\milli \bar}$ are required for optimum operation and the high-$P_0$ nozzle \cite{Ferrer2021} is perfectly suited. The jets ($\sim \SI{550}{\m \per \s}$) produced require a powerful, high-repetition-rate laser system ($10-\SI{15}{\kilo\hertz}$). For a more detailed explanation of the laser system see \cite{Kudryavtsev2016a,Verlinde2020}.
\subsection{Cooling of molecular degrees of freedom}
Up until now, characterization of the (translational) temperatures of the gas jets formed by the low-$P_0$ nozzle had only been performed with seeded \ch{^{63}Cu} (KU Leuven, \cite{Lantis2024, thesisClaessens}) or \ch{^{164}Dy} atoms (JGU Mainz, \cite{Munzberg2022}). For the low-$P_0$ nozzle, the jet is formed under flow match conditions with a stagnation pressure $P_0 = \SI{80}{\milli \bar}$ into a background $P_\text{bkg} = \SI{5.9e-3}{\milli \bar}$, giving an average translational temperature of $\SI{12.5 \pm 1.0}{\kelvin}$ over the full $\SI{60}{\mm}$ length with a room temperature gas cell. For a diatomic or polyatomic molecule, the vibrational and rotational degrees of freedom are cooled by low-energy broadside collisions with the cold carrier gas in the divergent part of the nozzle \cite{aquilanti1999}. In this short acceleration period of the gas jet, vibrational relaxation happens in general at a much lower rate than rotational relaxation, leading to three distinct molecular temperatures in the collision-free jet for the vibrational, rotational and translational degrees of freedom \cite{Smalley1977,Amirav1980,Kolodney1983}. For this reason, we performed a series of measurements to characterize the cooling properties of the low-$P_0$ hypersonic nozzle on the diatomic molecule $^{232}$ThO.  
\\
\\
Two high-lying electronic transitions, U-X \cite{Edvinsson1987} and B'-X \cite{Edvinsson1988} were probed. Combined with a second laser at $\SI{355}{\nm}$,the ThO molecules were ionized non-resonantly by excitation of an electron beyond the ionization potential \cite{Goncharov2005}. The first excitation step was provided by a dye laser (Credo, Sirah) or by a seeded single-mode pulsed dye amplifier (PDA) \cite{Verlinde2020} for broad-band or narrow-band spectroscopy, respectively. Seeding of the PDA was provided by a commercial external cavity diode laser (ECDL, Toptica) at $\SI{796}{\nm}$ for the U-X transition or by an optical parametric oscillator (OPO C-wave, H\"ubner photonics) at $\SI{660}{\nm}$ for the B'-X transition. Both PDA outputs were frequency doubled to execute the laser ionization schemes that are shown in Figs. \ref{fig:iglissetup_mainfigure} and \ref{fig:isotopologueshifts}. The thorium monoxide molecules for the temperature characterization were produced by ablating a commercially available metallic foil of $^{232}$Th (Goodfellows $99.5 \%$). Laser ablation was performed using the second harmonic of a Nd:YAG laser (INDI, Spectra Physics) at $\SI{532}{\nm}$, delivering $\SI{10}{\ns}$ pulses at a repetition rate of $\SI{20}{\Hz}$. Typical laser fluences on the metallic foil were found to be about $\SI{0.5}{\joule \per  \cm \squared}$.

\subsection{Vibrational temperature}
The vibration-translation equilibrium is often not achieved in supersonic expansion and the temperature associated with the vibration can be orders of magnitude higher than the translational or rotational temperature in the collision-free region. For the argon mass flow of our nozzle, one expects a freezing of the vibrational degree at around or slightly below $\SI{200}{\kelvin}$ when starting with room temperature ThO, based on a Landau-Teller model and accompanying experiments performed in \cite{McClelland1979}. Laser ablation studies of $^{232}$ThO found in the literature, however, report vibrational temperatures of $\SI{2500}{\kelvin}$, populating up to $16$ vibrational overtones in both argon \cite{Long2014} and helium free gas jet expansions \cite{Schmitz2019}. With the current IGLIS setup, significant cooling in the gas cell is however expected due to the relative longer extraction time of around $\SI{500}{\ms}$ \cite{thesisVerlinde} compared to these studies. 
\\
\\
With a broad-band dye laser (i.e. 2-3 GHz bandwidth), several regions with known transitions starting from the $\nu''=1$ vibrational state of the electronic ground state X could be covered. Not a single transition was observed in this study. At room temperature, only about $1.8\%$ of the thermal population would be in this $\nu''=1$ state, owing to the large spacing of $\SI{890.99}{\per \cm}$ between the $\nu''=0$ and $\nu''=1$ levels in the ground state. The most intense spectrum covering both a 0-0 and 1-1 vibrational transition is shown in Fig.~\ref{fig:iglissetup_mainfigure}. Here, B'-X has its 1-1 transition at $\SI{30243.7}{\per \cm}$ \cite{Goncharov2005}. Using a simple Boltzmann curve,
\begin{equation}
	\frac{N_{1}}{N_{0} + N_{1}} = e^{-\frac{E}{k_B T_v}}, \label{eq:BoltzmannVibr}
\end{equation}
an upper bound on the vibrational temperature ($T_v$) is estimated using the count rate $N_0$ of 0-0 and $N_1$ of 1-1. For $N_0$, the maximum of the resonance (Q-branch) is taken, for $N_1$, the $95\%$-interval ($2\sigma$) of the uncertainty on the background is taken. Assuming a Frank-Condon coefficient of 1, it is determined that $T_v < \SI{250}{\kelvin}$, indicating relaxation that is consistent with room temperature ThO at the nozzle entrance. While only an upper bound for $T_v$ is established, more effective cooling could be expected when higher $P_0$ is used \cite{McClelland1979} or in the case of polyatomic molecules \cite{Amirav1980}. Furthermore, the vibrational structure could be studied in our gas jet by simply using gas cells with shorter extraction times or by increasing intentionally the temperature of the gas $T_0$.
\subsection{Rotational temperature}
While estimates of the rotational temperature can be derived from the broad-band data, the measured intensities of the individual rotational transitions from narrow-band laser spectroscopy give direct access to the rotational temperature. 
Three measurements were performed with a transverse 2nd step to evaluate the evolution of the rotational temperature along the full gas jet: 
\begin{enumerate}
	\item All branches of the B'-X transition measured at the first few mm of the gas jet (exit plane) with 1st step laser in saturation and transverse,
	\item Q branch of U-X along the first $\SI{40}{\mm}$ with the 1st step laser in saturation and anti-propagating,
	\item Q branch of U-X along the full $\SI{60}{\mm}$ jet, with no powerbroadening and an anti-propagating 1st step laser.
\end{enumerate}
These allow for a consistency check between two different transitions and between individual branches. Additionally, the latter is a simultaneous measurement of both rotational and translational temperatures under maximal sensitivity conditions (see Sec.~\ref{sec:translationalTemp}). According to \cite{Herzberg1950}, the amplitude $A_J$ of a transition originating from a state $J$ with $v=0$ is given by  
\begin{equation} 
	A_J = \frac{2C\nu_J}{Z} S_J \exp{\left(-\frac{B_0J(J+1)hc}{k_B T_{\operatorname{rot}}}\right)}. \label{eq:line_intensities}
\end{equation} 
In the Boltzmann factor, $B_0 =\SI{0.331971}{\per \cm}$ is the ground state rotational constant for $v=0$ \cite{Dewberry2007} and $k_B/hc = \SI{0.6950348004}{\per\kelvin \per\cm}$ \cite{CODATA2018}. The multiplication factor consists of a constant $C$ related to the dipole transition strength, the transition frequency $\nu_J$ and the partition function $Z$. The angular momenta coupling is provided by the H\"onl-London factors $S_J$ \cite{Hansson2005}. For the two transitions in question, i.e., $\Delta \Lambda =1$, these are 
\begin{align}
	S_J^R &=  J+2, \notag \\
	S_J^Q &=  2J+1, \label{eq:HL} \\
	S_J^P &=  J-1. \notag
\end{align}
We extract here the rotational temperature using two methods. In the first method a Voigt function with a linear background is fitted to each clearly-separated resonance. By ignoring the $\nu$ dependence in Eq.~\ref{eq:line_intensities}, the fractional population $N_J$ of a state is first calculated from the experimentally determined $A_J$'s,
\begin{equation*}
	N_J = \frac{2(2J+1) A_J/S_J}{ \sum\limits_J 2(2J+1) A_J/S_J},
\end{equation*}
and related to the temperature via
\begin{equation*}
	N_J = (2J+1) \tilde{N_0}\exp{\left(-\frac{B_0J(J+1)hc}{k_B T_{\operatorname{rot}}}\right)}.
\end{equation*}
Here, $\tilde{N_0}$ is a fit parameter that denotes the model population in state $J = 0$ whereas $N_0$ is the experimentally determined population of $J = 0$ which is  accessible in the R-Branch using the R(0) line. A least squares fitting routine is then used to extract $T_{\operatorname{rot}}$.
\\
\\
In the case of the Q-branch of the B'-X, the individual peaks are merged but the branch as a whole is still clearly separated. Therefore, a second method was employed whereby a full branch of Voigt models with a linear background was fitted using least squares to the complete branch spectrum. The Lorentz contribution is a single parameter in the model with a fixed Gaussian contribution. The amplitudes are linked by
\begin{equation*}
	A_J =  \frac{A_2}{S_2}S_J \exp{\left(-\frac{B_0(J(J+1) - 6 )hc }{k_B T_{\operatorname{rot}}}\right)}
\end{equation*}
with the reasonable assumption that the difference in frequencies $\nu_J$ is negligible. $A_2$ was chosen as a common amplitude as it is found in all three branches. The rotational temperatures $T_{\operatorname{rot}}$ using both methods are listed in Table~\ref{tab:Trot_methods}.
\begin{table}[H]
	\caption{$T_{\operatorname{rot}}$ measured at different ranges in the gas jet: at the exit of the nozzle, over the first $\SI{40}{\mm}$ of the jet and over the full $\SI{60}{\mm}$ gas jet. The results of both methods are shown.}
    \centering
	\begin{tabular}{@{}lrrr@{}}
		\toprule
		\multicolumn{4}{c}{Rotational temperature $T_{\operatorname{rot}}$}            \\ \midrule
		\multicolumn{1}{c}{Range jet}   & \multicolumn{1}{c}{Transition}  & \multicolumn{1}{c}{Method 1 [K]}  & \multicolumn{1}{c}{Method 2 [K]}\\ \midrule
		exit	 			& B'-X R(J)   &   $7.0(5)$   & $6.8(2)$    \\
		& B'-X Q(J)   &   ---   			& $7.3(2)$    \\
		& B'-X P(J)   &   $6.4(4)$   &  $7.0(2)$   \\ \midrule
		$0-\SI{40}{\mm}$ 	& U-X Q(J)   &   $7.1(1)$   & $6.8(2)$    \\  \midrule
		$0-\SI{60}{\mm}$ 	& U-X Q(J)   &   $7.2(6)$   & $6.6(3)$    \\
		\bottomrule
	\end{tabular}
	\label{tab:Trot_methods}
\end{table}
Both methods give consistent temperatures, but method 2 provides a more precise value owing to its constricted model. The standard deviation of the temperatures seen in method 2 for all three branches of B' is $\SI{0.2}{\kelvin}$, confirming the validity of the error provided by the fit.  The weighted average temperature at the nozzle exit is $\SI{7.1 \pm 0.1}{\kelvin}$. This temperature is maintained along the jet, indicating good homogeneity. To account for instabilities from ablation/excitation laser observed in the $\SI{60}{\mm}$ spectrum, the more conservative $T_{\operatorname{rot}} = \SI{7.2 \pm 0.6}{\kelvin}$ from method 1 is chosen as the final temperature.

\subsection{Translational temperature} \label{sec:translationalTemp}
\begin{figure}[H]
	\includegraphics{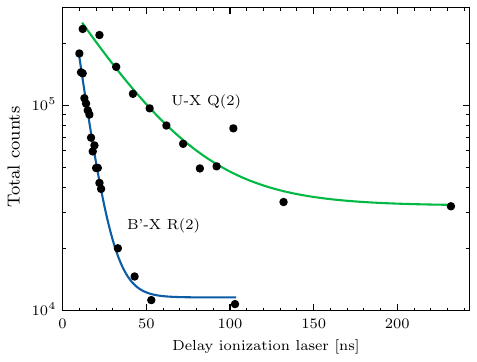}
	\caption{Measured half-lives of two well-resolved rotational states that are used for the translational temperature determination. For the B' state the R(2) transition was chosen which measures $J'=3$ as the more collapsed Q-branch has overlapping peaks. In both cases, an exponential function with a constant background was fitted. Different measurement times and laser powers were used for each transition, resulting in the different backgrounds.}
	\label{fig:halflifes}
\end{figure}

The overall cooling of the velocity components is determined by the Mach number $M$ which can be expressed as  
\begin{align}
	M = \sqrt{\frac{2}{\gamma -1} \left(\frac{T_0}{T_t} -1\right)}.
\end{align}
Using the adiabatic constant $\gamma =5/3$ for argon gas and measuring the  stagnation temperature $T_0$ (i.e. temperature of the gas in the cell) and translational temperature in-gas-jet $T_t$, the Mach number can be obtained. As contraction and expansion cycles along the gas jet lead to oscillations in $M$, we determine $M$ both at the nozzle exit and over the full jet using the B'-X and U-X transitions respectively. In the latter, a single Q-branch measurement allowed the determination of both rotational and translational temperatures simultaneously over the full gas jet. The use of resonance ionization spectroscopy to determine the translational jet temperature is outlined here and more details can be found in \cite{Ferrer2021}. In short, the Doppler broadening and thus temperature found in the linewidth of the peaks can be isolated provided that power broadening is negligible. In order to achieve this, optical density filters were used on the first (excitation) laser step until the linewidth stopped decreasing. The second step (ionization) laser was also delayed to avoid overlap with the first and thus contributing to the power broadening. In this case the linewidth is fully determined by the Doppler broadening $\Delta \nu_\text{Doppler}$, laser linewidth $\Delta \nu_\text{laser}$, natural linewidth $\Gamma$ and collisional broadening. The latter is only a few MHz for jets from the low-$P_0$ nozzle and can be ignored. The full width at half maximum (FWHM) $\Delta \nu$ is \cite{Olivero1977}
 \begin{align}
\Delta \nu \approx 0.5346 \Gamma  + \sqrt{0.2166 \Gamma^2 + \Delta \nu_\text{laser}^2 + \Delta \nu_\text{Doppler}^2 }.
\end{align}
 Both Gaussian contributions to the linewidth can be conveniently expressed in MHz as 
\begin{align}
	\Delta \nu_{\operatorname{Laser}} = \frac{441}{\Delta \tau }, \quad \Delta \nu_{\operatorname{Doppler}} = 7.16 \cdot 10^{-7} \nu_0 \sqrt{\frac{T_t}{A}}
\end{align} 
where $\Delta \tau$ is the PDA pulse-width in ns, $\nu_0$ is the center frequency of the transition, $A$ is the mass of the species in atomic mass units and $T_t$ is the (translational) temperature in kelvin. The pulse-widths (FWHM) of the PDA output are $\SI{2.82\pm 0.07}{\ns}$ and $\SI{3.23\pm 0.08}{\ns}$ for the respective U-X and B'-X transitions. The natural linewidth is related to the Einstein coefficient $A_{\operatorname{Ein}}$ as $\Gamma = A_{\operatorname{Ein}}/(2 \pi)$ and is extracted from the half-lives of the excited J-states of the electronic U and B' states. This was done by delaying the second step laser with respect to the first one and fitting an exponential function to the observed count rates. The results are shown in Fig.~\ref{fig:halflifes}. The half-life is measured to be $\SI{5.1 \pm 0.2}{\ns}$ and $\SI{23 \pm 3}{\ns}$ for the B' $J'=3$ and U $J'=2$ states respectively. For the case of U-X, we can rescale the half-life using Eq.~\ref{eq:HL} to get a half-life of $\SI{41\pm 5}{\nano \s}$ for $J'=4$. This is in agreement with a previous broadband measurement in the literature that was performed on the unresolved Q-branch head where a half-life of $\SI{39.5\pm 1.4}{\nano \s}$ was measured \cite{Kokkin2014}. Based on the $\SI{20}{\kelvin}$ rotational temperature, it is inferred from those measurements that Q(4) would have been the most intense transition in the results in \cite{Kokkin2014}.
 A stagnation temperature $T_0$ of $\SI{305(2)}{\kelvin}$ is measured through the Doppler shift $\delta \nu_{\text{Doppler}}$ when the laser beam counter-propagates to the gas jet (see Fig.~\ref{fig:iglissetup_mainfigure} $\mathbf{a}$),
\begin{align}
\delta \nu_{\text{Doppler}} =  - \frac{\sqrt{\frac{2 \gamma k_\text{B} (T_0 - T_t)}{m_{\ch{Ar}} (\gamma -1 )}}}	{\lambda_{\operatorname{vac}} \times 10^6}.
\end{align}
where $\lambda_{\operatorname{vac}}$ is the non-shifted wavelength in vacuum given in nm and $m_{\ch{Ar}}$ is the mass of natural abundant argon. The extracted translational temperature $T_t$ is $\SI{8 \pm 2}{\kelvin}$, corresponding to $M = 10.6(14)$, at the nozzle exit and  $\SI{12.5 \pm 2.5}{\kelvin}$ with $M = 8.4(9)$ over the full $\SI{60}{\mm}$ gas jet. This is consistent with the flow mapping conducted with $^{63}$Cu in \cite{Lantis2024,thesisClaessens} but with a larger uncertainty as the probing laser has roughly the same linewidth as the Doppler contribution in the case of ThO.
\section{Target production} \label{sec:targets}
Based on the work of \cite{Stricker2025}, ablation sources that inherently contain the ThO were chosen for spectroscopy. The main requirements for the $^{229,230}$ThO targets were the consistent delivery of neutral molecules and chemical stability of the target during the ablation process. Triflate-based targets \cite{Artes2025} were chosen which are normally used in ion beam applications in large accelerator facilities such as GSI Helmholtzzentrum für Schwerionenforschung, Germany, or the Grand Acc\'el\'erateur National d’Ions Lourds (GANIL), France. In such ion beam experiments, these targets are exposed to large stress over a longer period of time~\cite{Dmitriev2015,Dullmann2023}, which should make them ideal for laser ablation experiments.
\\
\\
 Triflic acid is used to produce thoriumtriflate (ThOTf$_4$) for the production of electrochemically deposited thorium oxide laser ablation targets. Two $^{230,232}$ThOTf$_4$ and one $^{229,232}$ThOTf$_4$ solutions were prepared. Each $^{230,232}$Th solution contained \SI{7}{mg} ($^{232}$Th) thorium nitrate pentahydrate (\SI{12}{Bq}) and \SI{64}{\micro g} of ($^{230}$Th) thorium chloride (\SI{30}{kBq}) in \SI{10}{mL} of MilliQ water. Similarly, the $^{229,232}$Th solutions contained \SI{7}{mg} ($^{232}$Th) thorium nitrate pentahydrate (\SI{12}{Bq}) and \SI{6}{\micro g} of ($^{229}$Th) thorium chloride (\SI{44}{kBq}) in \SI{10}{mL} of MilliQ water. The solutions were heated with a small stoichiometric excess of triflic acid, and the water was then evaporated to dryness. The $^{230,232}$Th residues were dissolved in N,N-dimethylformamide (DMF) and deposited galvanostatically on titanium foils in an electrochemistry cell~\cite{Artes2025}. In the case of $^{229,232}$Th, a mixture 9:1 of iso-propanol and iso-butanol (v/v) was used instead, and this solution was used for both $^{229}$Th-containing targets. During the electrochemical deposition process, the thorium triflate was completely converted to thorium oxide. The specifications of the prepared targets are listed in Table~\ref{tab:target_properties}.
\begin{table}[htbp]
	\centering
	\caption{Properties of the prepared thorium triflate targets. Mass and activity are given for the sum of both radioisotopes. For the laser spectroscopy studies, only two targets were used: a $^{230,232}\text{ThO}$ target with an activity of $\SI{20}{\kilo \becquerel}$ and a $^{229,232}\text{ThO}$ target with $\SI{11}{\kilo \becquerel}$. }
	\label{tab:target_properties}
	\begin{tabular}{@{}l c c c c @{}}
		\toprule
		Isotopes & Mass (\unit{mg}) & A (\unit{kBq}) & Dia. (\unit{cm}) & Thickness (\unit{\micro\meter}) \\
		\midrule
		$^{230/232}\text{Th}$ & 1.95 & 20 & 2.5 & $1.8 \pm 0.4$ \\
		$^{230/232}\text{Th}$ & 1.32 & 19 & 2.5 & $2.0 \pm 0.4$ \\
		$^{229/232}\text{Th}$ & 2.67 & 11 & 1.0 & $14.1 \pm 0.2$\\    
		$^{229/232}\text{Th}$ & 0.34 & 5  & 1.0 & $18.5 \pm 0.2$ \\
		\bottomrule
	\end{tabular}
\end{table}

The activity was quantified by measuring the characteristic alpha spectrum, and the thickness of the samples was determined using a laser microscope (KEYENCE VK-X3000). In addition, scanning electron (SE) and backscattering electron (BSE) microscopy were used to determine the morphology using a Philips XL30. The distribution of activity was determined by radiography (FUJIFILM FLA 7000) (Fig.~\ref{fig:isotopologueshifts}$\mathbf{a}$), which shows that the activity is distributed evenly across the entire target.
\\
\\
For each isotopologue measurement, one triflate target was installed together with the metallic foil of natural thorium for reference. Both were attached to a copper backing plate to improve thermal and electric conduction via screws. The triflate source was further glued to the copper with silver epoxy. Unfortunately, the unsuppressed ablation background did not show any identifiable mass peaks, making the ratio between Th and ThO ions impossible to estimate. An accurate total efficiency could not yet be ascertained but is expected to be low due to the inherent nature of the ablation process. Nevertheless, optimization of the IGLIS technique over the last decade has recently enabled in-gas-jet laser spectroscopy measurements of $^{254 \text{m}}$No with $0.1 \%$ overall efficiency \cite{thesisIvandikov}.

\section{Results}\label{sec:results}
\begin{figure*}
	\includegraphics{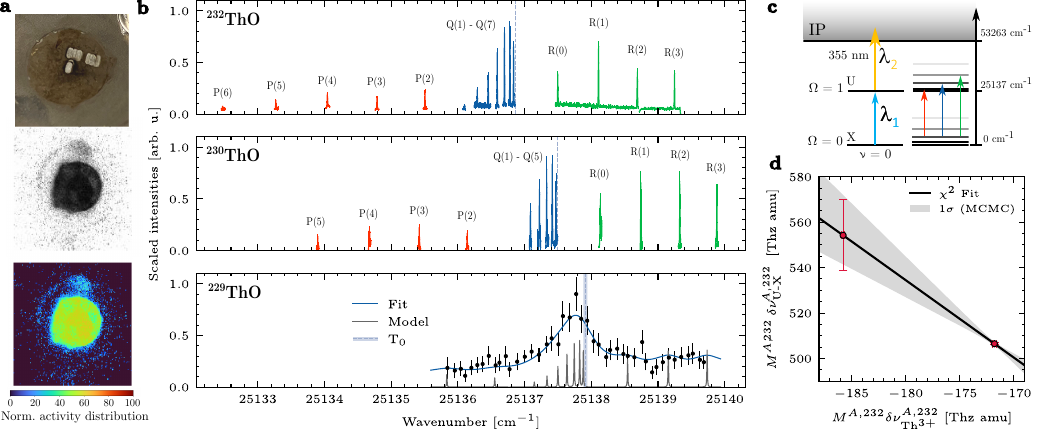}
	\caption{Photographic (Top) and radiographic images (Middle and Bottom) of one of the ablation targets is shown in $\mathbf{a}$. The etched rectangular sections on the photograph (Top) are regions that are ablated. This particular target contains $\SI{11}{\kilo\becquerel}$ of $^{229}$Th. The measured spectra using the U-X transition ($\mathbf{b}$) for the three isotopologues are color coded according to their branch and each scan has been rescaled to follow the intensities of the $\SI{7.2}{\kelvin}$ temperature distribution. The found term energies $T_0$ and the spectra themselves are not Doppler corrected in $\mathbf{b}$. The $^{229}$ThO data are fitted by the model in gray, where the FWHM is reduced to $\SI{250}{\MHz}$ to show the individual contributions. Laser ionization scheme is shown to the right ($\mathbf{c}$). The molecular King plot is shown bottom right ($\mathbf{d}$) and the fit from the $\chi^2$ routine and uncertainty band of the MCMC is depicted.}
	\label{fig:isotopologueshifts}
\end{figure*}

Rotational spectra were measured for all three isotopologues using the U-X transition and are shown in Fig.~\ref{fig:isotopologueshifts}. These measurements were conducted with the first laser step being aligned anti-propagating with the direction of flow and the second step transverse to the flow (cf. Fig.~\ref{fig:iglissetup_mainfigure}). Hence, the resulting spectra are Doppler red-shifted by $\SI{1.40 \pm 0.01}{\giga \Hz}$. The narrow-band spectra for both $^{232}$ThO and $^{230}$ThO were taken with spectral resolutions around $200-\SI{250}{\mega \hertz}$ FWHM. In the case of $^{229}$ThO, only broad-band spectra were taken due to the low count rate ($\sim 2$~cts/s on resonance). Besides U-X, the B'-X transition was also measured but only for $^{232}$ThO (see Fig.~\ref{fig:iglissetup_mainfigure} $\mathbf{e}$). Both lasers were overlapped in this case, in a transverse geometry close to the nozzle with high power, resulting in a resolution of $\SI{500}{\mega \hertz}$. The typical laser powers used for narrow-band spectroscopy are around $\SI{5}{\milli \watt}$ (high) for the first laser step which is attenuated by an optical density filter with a factor 500-1000 for high-resolution spectroscopy. The non-resonant laser step operated at $\SI{25}{\watt}$, running at a $\SI{10}{\kilo \hertz}$ repetition rate.
 \\
 \\
 For both transitions, the first harmonic of the PDA was recorded using a HeNe-calibrated Highfinesse WS7/60 wavelength meter. In the range of $\SI{800}{\nm}$ (U-X) and $\SI{650}{\nm}$ (B'-X), the relative frequency reproducibility is within $\pm \SI{5}{\mega \hertz}$ and $\pm \SI{12}{\mega \hertz}$ respectively and the absolute accuracy is $\pm \SI{60}{\mega \hertz}$ \cite{Verlinde2020a}. A dynamic Stark shift of about $\SI{80}{\mega \hertz}$ was observed in B'-X between low-power and high-power spectra. To account for any residual Stark shift, a systematic uncertainty of $\SI{40}{\mega \hertz}$ is added to the absolute accuracy.  In the case of U-X, no systematic shifts were observed and the peak centers were within $\pm \SI{5}{\mega \hertz}$.

\subsection{Molecular structure}
 Each individual rotational state was measured and its spectra fitted by a Voigt line-shape on a linear background using a $\chi^2$ routine. A model based on a Hamiltonian which incorporates $\Omega$-doubling relates the found peak positions to the diatomic constants.  A rotational level with quantum number J has its energy at
\begin{align} 
	E(J) = T_0 + \left(B_0 \pm \frac{q}{2}\right) J(J+1)
	\label{eq:molecularEnergyLevels}
\end{align} 
where $T_0$ is the electronic term energy, $B_0$ the rotational constant at the $v=0$ vibrational level and $q$ the $\Omega$-doubling term. For the ground state term X$^1\Sigma^+$, both $T_0$ and $q$ are zero. Both excited terms are subject to $\Omega$-doubling that splits the e/f symmetry levels with the e-levels being higher in energy (i.e. $q>0$). Fitting using this Hamiltonian was conducted by the program PGopher \cite{Western2017}. Higher-order distortion effects were not taken into account as they resulted in unrealistically high values.  The diatomic constants for the two isotopologues are listed in Table~\ref{tab:rotational_parameters}.
\\
\\
For both the ground state and the U state, the measured change in rotational constants between the two isotopologues is a pure mass effect, i.e $B_0^{232} \mu^{232} = B_0^{230} \mu^{230}$ within experimental uncertainties. Here, $\mu$ denotes the reduced mass of the thorium and oxygen atoms. For the ground state, the difference between the mass scaled and the measured $B_0$ of $^{230}$ThO is $\SI{-5(5)e-5}{\per \cm}$ while for the excited U state it is $\SI{-2(6)e-5}{\per \cm}$. In the case of $^{230}$ThO, ablation production decreased over time and was overall less stable than in the case of the metallic $^{232}$Th target. This resulted in decreased precision on the $^{230}$ThO states which is reflected in its molecular constants. Both B'-X and U-X transitions provide a similar $B_0$ for the ground state which is, together with the B' molecular constants, consistent with literature values.
\\
\\
 Our rotational constant for the excited state U disagrees significantly with the value derived from both U-H ($J \geq 4$) and U-X ($J \geq 12$) transitions using spectrography \cite{Edvinsson1987}. The reason for this is most likely the non-local perturbation present in this term as was noted in this previous work. Therefore the stated $B_0$ and $q$ can only be regarded as effective parameters for the measured range of J values. Combination defect analysis confirms the presence of the perturbation as the $\Omega$-doubling grows non-linearly over $J(J+1)$ even over the limited set of lines measured here. 
\begin{table}[htbp]
	\centering
	\caption{Experimental diatomic parameters in \si{\per \cm} for the different states of $^{230,232}$ThO and the literature values for $^{232}$ThO. All $T_0$ constants are corrected for the Doppler and Stark shifts and have an additional systematic error (in square brackets) from the absolute wavelength meter accuracy ($\SI{60}{\MHz}$) and Stark shift ($\SI{40}{\MHz}$).}
	\label{tab:rotational_parameters}
	\begin{tabular}{@{} l c c c @{}}
		\toprule
		Constant & $^{232}\text{ThO}$ & $^{230}\text{ThO}$ & Lit. \cite{Dewberry2007,Edvinsson1988,Edvinsson1987} \\
		\midrule
		\multicolumn{4}{@{}l}{\textit{X\,${}^1\Sigma^+$}} \\
		$B_0$ & $0.33197(1)$ & $0.33221(5)$ & $0.3319710(1)$ \\
		
		\addlinespace
		\multicolumn{4}{@{}l}{\textit{B$'\,(\Omega = 1)$}} \\
		
		$T_0$    & $30312.9827(6)[33]$ & --- & $30312.99(1)$ \\
		$B_0$    & $0.32664(1)$        & --- & $0.32666(2)$  \\
		$10^5 q$ & $14(1)$             & --- & $16(4)$       \\
		
		\addlinespace
		\multicolumn{4}{@{}l}{\textit{U\,$(\Omega = 1)$}} \\
		$T_0$    & $25136.9090(3)[20]$ & $25137.5436(5)[20]$ & $25136.90(1)$ \\
		$B_0$    & $0.31837(2)$        & $0.31857(6)$        & $0.31875(2)$  \\
		$10^5 q$ & $5(1)$              & $4(4)$              & ---           \\
		\bottomrule
	\end{tabular}
\end{table}

\subsection{Isotopologues shifts}
The isotopologue shift of heavy diatomic molecules such as ThO can be accurately approximated using the linearized isotope-shift equation described in \cite{Athanasakiskaklamanakis2023}, 
\begin{align}
	\delta \nu^{A,232} = \Psi_0 \delta \left \langle r^2 \right \rangle_\text{Th}^{A,232} + \Xi_0 \frac{1}{M_\text{Th}^{A,232}}.
\end{align}  
Here, $\Psi_0$ and $\Xi_0$ act as the field and mass shift factors respectively, similarly as in atomic isotope shifts. The isotopologue shift with reference $^{232}$ThO is taken with respect to the state term energies such that $\delta \nu^{A,232} = T^A_0 - T^{232}_0$ and the mass coefficient $M^{A,232}$ is equal to $m_\text{Th}^{232}m_\text{Th}^{A}/(m_\text{Th}^{232}-m_\text{Th}^{A})$ with $m$ the atomic mass. Due to the low count rate, the spectrum of $^{229}$ThO was only recorded using broadband lasers with no individual rotational or hyperfine peaks being visible. 
\\
\\
Using the known molecular constants of $^{232}$ThO, a diatomic model using the level structure provided by Eq.~\ref{eq:molecularEnergyLevels} (with $q=0$) and intensities by Eq.~\ref{eq:line_intensities} was constructed using the SATLAS2 package in python \cite{Gins2024}. Every individual peak was linked to a common linewidth. This model was fitted to the spectrum on top of a linear background were both $T_0$ and the common linewidth were simultaneously varied. The underlying hyperfine structure was not accounted for. The model was tested on broadband measurements of $^{230,232}$ThO with a similar step size of about $\SI{0.1}{\per \cm}$ as the $^{229}$ThO spectrum. The term energy of the broad-band measurements were within $\SI{0.014}{\per \cm}$ of the narrow-band spectra. A Doppler-corrected $T^{229}_0 = \SI{25137.96(3)}{\per \cm}$ was found from the fit with a common Voigt linewidth of $\SI{16(3)}{\giga \hertz}$ .
\\
\\
To extract $\Psi_0$ and $\Xi_0$, a molecular King plot analysis was performed which compared the molecular shifts to shift measured for ionic Th$^{3+}$ in \cite{Zitzer2025} and their shift factors calculated in \cite{Chakraborty2026}. The ionic factors $F,K$ and isotope shifts $\delta \nu^{A,232}_{\text{Th}^{3+}}$ allow us to extract the molecular factors via the linear relationship
\begin{align}
M^{A,232} \delta \nu^{A,232}_{\text{ThO}} = \Xi - \frac{\Psi_0}{F}K + \frac{\Psi_0}{F} M^{A,232} \delta \nu^{A,232}_{\text{Th}^{3+}}.
\end{align}
The King plot is shown in Fig.~\ref{fig:isotopologueshifts}($\mathbf{d})$ where the slope of the linear fit is $\Psi_0/F$ and the intercept is $\Xi - K\Psi_0/F$. As only two shifts were measured, the parameter space was explored using a Markov Chain Monte Carlo (MCMC) algorithm to estimate the uncertainty. The mean parameter values were derived from a simple least squares fit. We find a field shift factor of $\Psi_0 = -\SI{4.0\pm 1.3}{\cm^{-1} /fm^2}$  and a mass shift factor of $\Xi_0 = (-2 \pm 6) \times 10^3 \text{ cm$^{-1}$ amu}$ for the U-X transition. Change in reference isotopologue or using a ladder technique for the shifts did not improve the uncertainty. The molecular field shift factor is 3 times larger than that of its constituent Th$^{2+}$ ion, and 1.4 to 3 larger than those of the Th$^{+}$ and Th$^{3+}$ ions, respectively \cite{Chakraborty2026,Campbell2011,Zitzer2025,Safronova2018a}.

\section{Molecular structure calculations}\label{sec:theory}
\begin{figure}
	\includegraphics[width=.5\textwidth]{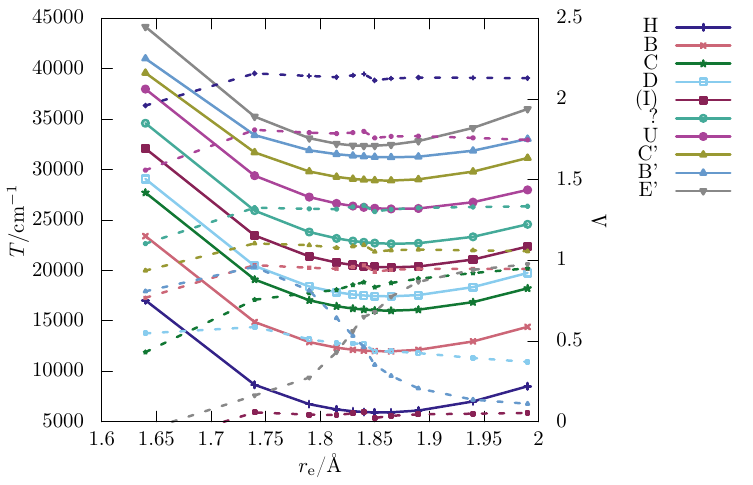}
	\caption{Potential energy curves of the energetically lowest $\Omega=1$ states (solid lines) and effective electronic orbital angular momentum projections on the molecular axis $\Lambda$ as a function of the bond length $r_\mathrm{e}$ (dotted lines) computed at the level IH-FSCCSD(0,2)/cv3z (3z). Lines are shown to guide the eye.}
	\label{fig:potential_curves}
\end{figure}

Transition energies of ThO were calculated at the level of relativistic
Fock-Space Coupled Cluster including Singles and Doubles amplitudes (FSCCSD).
We employed the intermediate Hamiltonian formalism \cite{landau:2001} (IH-FSCCSD) to mitigate
intruder states and improve convergence properties of the FSCCSD calculations.  All calculations were carried out 
with the program package DIRAC26~\cite{DIRAC26}. We employed correlation-consistent all-electron basis sets developed by Dyall, including core-valence correlation functions, of double-, triple-, and quadruple-$\zeta$-quality (cv2z, cv3z, and cv4z, respectively) \cite{dyall:2002,dyall:2006}. In order to describe higher-lying electronic states accurately we doubly augmented the basis set on Th with diffuse functions.  In the closed-shell reference, we correlated 34 electrons,  corresponding to the atomic configurations $\mathrm{Th6p^65d^{10}6s^25p^6}$ and $\mathrm{O2p^62s^2}$. We correlated all spinors in the energy range of $-12.5\,E_\mathrm{h}$ to $300\,E_\mathrm{h}$, i.e., 612 inactive virtual spinors for the cv3z basis set (denoted 3z) and 400 for the cv2z basis set (denoted 2z). For estimating the uncertainty due to the basis set and the chosen correlated space, we used a smaller correlated space in the energy range of $-5\,E_\mathrm{h}$ to $5\,E_\mathrm{h}$ and compared results with the cv3z basis set (26 inactive occupied and 418 inactive virtual spinors correlated) with results obtained with the larger cv4z basis set (26 inactive occupied and 584 inactive virtual spinors correlated). We denote these calculations in the following 3zs and 4zs, respectively.
\\
\\
We followed
Ref.\,\cite{tecmer:2018} for selecting the active model space in the 0-2 sector (0
holes, 2 electrons) from the Th-centered $\mathrm{7s}$, $\mathrm{6d}$ and $\mathrm{5f}$ shells, comprising 26 spinors in total. To facilitate convergence, we treated the $\mathrm{5f}$ shells (14 spinors) in the intermediate model space (energy range $-0.3\,E_\mathrm{h}$ to $-0.2\,E_\mathrm{h}$), leaving 12 spinors associated with the  $\mathrm{7s}$ and $\mathrm{6d}$ shells. To assess the sensitivity to the choice of intermediate model space, we performed additional calculations in which the Th-centered $\mathrm{7p}$ spinors were included in the intermediate model space. These calculations were performed with the cv3z basis set and are denoted 3zl. 
\begin{table*}
\caption{Computed spectroscopic parameters of the electronic ground state ($\Omega=0$) and the ten energetically lowest $\Omega=1$ states dominated by $6d^2$ configurations localised mainly at Th at the level of IH-FSCCSD(0,2)/cv3z (3z). States are named according to their energetic ordering in the IH-FSCCSD(0,2)/cv3z calculation by roman numerals. Assignments to spectroscopic labels are provided according to matching of experimentally observed energetic orders and characterizations of the calculated state. In cases the assignment is uncertain a ? is added. State (xix) has not been observed in experiment (see also the discussion in Ref. \cite{zaitsevskii:2025}).  Excitation wavenumbers are provided for the vibrationally corrected  $0\leftarrow0$ transition and rotational constants of ${}^{232}$ThO, $B_0$, are given for the vibrational ground state in $\mathrm{cm}^{-1}$. Approximate electronic orbital angular momentum projections on the molecular axis $\Lambda$ and dominant configurations are provided for a bond length of $1.84\,$\AA. The assignment follows the energetic ordering described in Refs. \cite{Goncharov2005}, except for the B' and E' states, whose ordering depends strongly on the chosen active space (see the discussion in the text).}
\label{tab:states}
\begin{tabular}{llS[table-format=5.0]S[table-format=1.4]S l} 
\toprule
{State} & Spectr. Label &  {$T_\mathrm{0}$} & {$B_\mathrm{0}$} & {$\Lambda$} & {Dominant config. at $r=1.84\,$\AA} \\
\midrule
{(0)}&{ X }& {--}& 0.3314 & 0 & $97\,\%\,{}^1\Sigma_{0+}\mathrm{7s_{1/2}^2}$\\
{(i)}&{ H }& 5908 & 0.3261 & 2.1  & $99\,\%\,{}^3\Delta_{1}\mathrm{7s_{1/2,-1/2}6d_{3/2,3/2}}$\\
{(ii)}&{ B }& 11962& 0.3247 & 0.94 & $77\,\%\,{}^3\Pi_{1}\mathrm{7s_{1/2,+1/2}6d_{3/2,+1/2}}$\\
{(iii)}&{ C }& 15997& 0.3240 & 0.86 & $61\,\%,{}^3\Pi_{1}\mathrm{7s_{1/2,-1/2}6d_{5/2,+3/2}}$\\
{(iv)}&{ D }& 17422& 0.3247 & 0.44 & $75\,\%\,{}^3\Sigma_{1}\mathrm{7s_{1/2,+1/2}6d_{5/2,+1/2}}$\\
{(v)}&{ ? }& 20291& 0.3233 & 0.05 & $95\,\%\,{}^3\Sigma_{1}\mathrm{6d_{3/2,-3/2}6d_{5/2,+5/2}}$ \\
{(vi)}&{ I?  }& 22626& 0.3224 & 1.3  & $62\,\%,{}^1\Pi_{1}\mathrm{6d_{3/2,-1/2}6d_{3/2,+3/2}}$\\
{(vii)}&{ U }& 26077& 0.3222 & 1.8& $82\,\%,{}^3\Delta_{1}\mathrm{6d_{3/2,+3/2}6d_{5/2,-1/2}}$\\
{(viii)}&{ C' }& 28881& 0.3240 & 1.1  & $65\,\%,{}^1\Pi_{1}\mathrm{6d_{3/2,-3/2}6d_{5/2,+5/2}}$\\
{(x)}&{ E'? }& 32409 & 0.3304 & 0.65 & \parbox[t]{5cm}{$44\,\%,{}^3\Sigma_{1}\mathrm{6d_{5/2,+3/2}6d_{3/2,-1/2}}$ with large 7s5f admixture}\\
{(ix)}&{ B'? }& 31124 & 0.3234 & 0.28 & $70\,\%,{}^3\Sigma_{1}\mathrm{6d_{5/2,+3/2}6d_{3/2,-1/2}}$\\
\bottomrule
\end{tabular}
\end{table*}
The nuclear charge density of nucleus $A$ was 
described by a normalized Gaussian function $\rho_A(\vec{r}_i)=\rho_0
\mathrm{e}^{-\zeta \left|\vec{r}_i-\vec{r}_A\right|^2}$ with $\zeta=
\frac{3}{2\langle r^2 \rangle}$ being characterized by an empirically
determined root-mean-square nuclear radius of $\langle r^2 \rangle^{1/2} =
\left[0.836 A^{1/3} + 0.570\right]\,\mathrm{fm} $ \cite{visscher:1997a}. Here $\vec{r}_i$ and 
$\vec{r}_A$ denote the position vector of electron $i$ and nucleus $A$,
respectively. Calculations with the 2z and 3z configurations were performed for bond lengths of $r_{\mathrm{ThO}} =$ 1.64, 1.74, 1.79, 1.815, 1.83, 1.84, 1.85, 1.865, 1.89, 1.94,1.99~$\AA$,
which covers the region around the equilibrium internuclear distance of the electronic
ground state, 1.84\,$\AA$. For each bond length, the isotope shift constant
$\Psi(r_{\mathrm{ThO}})=\frac{\partial\delta
\widetilde{\nu}}{\partial\delta\langle r^2 \rangle}(r_{\mathrm{ThO}})$ is
determined by linear regression of seven excitation energies at values
$\langle r^2 \rangle =$\,30.4846, 31.4846, 31.9846, 32.5689, 32.9846, 33.4846, 34.4846$\,\mathrm{fm}^2$ of the Th nuclear radius.
\\
\\
 To characterize the electronic states beyond their energetic ordering and $\Omega$,
we analyzed the contributions of molecular spinors to individual excitations and calculated the expectation value
of the orbital angular momentum projection along the molecular axis $\Lambda=\langle\hat{L}_z\rangle$. Although $\Lambda$ is not an exact quantum number in the relativistic calculations, it provides an approximate characterization of the spin and orbital angular momentum character of the states, reflecting their composition in terms of $\Lambda$ and $\Sigma$ in the absence of spin-orbit coupling. The percentages assigned to the dominant configurations are calculated from squared wave function amplitudes. The corresponding mean-field configurations are characterized by leading atomic contributions to the molecular spinors determined using Mulliken population analysis. The excitation energies corresponding to $\Omega=1$ are listed alongside approximate $\Lambda$ values for different basis sets in Table \ref{tab:states}. We note that the energetic ordering of the (ix) and (x) states, corresponding to E' and B', is reversed for 2z and 3zl calculations compared to the other calculations. Both states exhibit a strong mixture of similar configurations, which depends on the bond length. The bond length dependence of $\Lambda$ hints towards an avoided crossing of these two electronic states around $1.84\,\mathrm{\AA}$ (see Fig.~\ref{fig:potential_curves}). The possible avoided crossing suggests that non-adiabatic effects may be relevant for the B' and E' states, although their magnitude is not quantified in the present work. We, therefore, cannot provide a certain assignment for these two states (see Fig.~\ref{fig:potential_curves}). We tentatively assign the E' and B' states according to calculations with a larger intermediate Fock space at 1.84\,\AA\ (3zl). The computed field shift parameters are shown for the 10 energetically lowest states in Table~\ref{tab:fieldshift}. 
\begin{table*}[htb]
	\caption{Molecular field shift constants at a fixed bond length of 1.84\,\AA, $\Psi(1.84\,\mathrm{\AA})$, and for the vibronic 0-0 transition including vibrational corrections, $\Psi_0$. States are labeled by roman numerals according to the energetic ordering in the respective calculation. Note that the energetic ordering of the B' and E' states (labeled (ix) and (x) according to energetic order) is reversed in the 2z and 3zl calculations compared with the 3z, 3zs, and 4zs calculations. Numerical uncertainties are provided in parentheses. All calculations are computed at the level of relativistic IHFSCCSD(0,2) with different basis sets and active space configurations as described in the text.}
	\label{tab:fieldshift}
    \begin{tabular}{lS[table-format=-1.2(1)]S[table-format=-1.2(1)]S[table-format=-1.3(1)]S[table-format=-1.3(1)]S[table-format=-1.2(1)]S[table-format=-1.2(2)]}
    \toprule
    State & \multicolumn{5}{c}{$\Psi(1.84\,\mathrm{\AA})\,\mathrm{cm}^{-1}/\mathrm{fm}^2$}
    & {$\Psi_0\,\mathrm{cm}^{-1}/\mathrm{fm}^2$} \\
    \cmidrule(lr){2-6}
    {} & {2z} & {3z} & {3zl} & {3zs} & {4zs} & {3z} \\
    \midrule
    {(i)}   & -1.61(4) & -1.60(6) & -1.621(4) & -1.590(4) & -1.595(5) & -1.58(3) \\
    {(ii)}   & -1.66(4) & -1.70(6) & -1.747(4) & -1.679(4) & -1.668(5) & -1.68(3) \\
    {(iii)}   & -1.91(4) & -1.95(6) & -1.993(5) & -1.958(5) & -1.935(5) & -1.94(5) \\
    {(iv)}   & -1.92(4) & -1.97(6) & -1.973(5) & -1.933(5) & -1.919(5) & -1.94(5) \\
    {(v)} & -3.17(4) & -3.19(6) & -3.195(7) & -3.169(8) & -3.175(8) & -3.17(13) \\
    {(vi)}   & -3.08(4) & -3.15(6) & -3.083(7) & -3.118(8) & -3.106(8) & -3.14(15) \\
    {(vii)}   & -3.34(4) & -3.41(6) & -3.395(8) & -3.378(8) & -3.370(8) & -3.24(32) \\
    {(viii)}  & -3.06(4) & -3.15(8) & -3.135(7) & -3.125(7) & -3.114(8) & -3.15(22) \\
    {(ix)} & -2.32(4)& -2.94(6) & -2.800(6) & -2.774(6) & -2.621(7) & -3.07(60) \\
    {(x)} & -3.09(4)  & -2.55(6) & -3.168(7) & -2.657(7) & -2.780(7) & -2.52(60) \\
    \bottomrule
\end{tabular}
\end{table*}
The uncertainty of the present calculations is assumed to be composed of independent uncertainties stemming from the incompleteness of the basis set, active correlated space and Fock space selections, higher-order coupled cluster excitations and vibrational effects. We focus this discussion on the used experimental transitions, U-X and the mixture B-X and E'-X. From the comparison of the c3z and c4z basis sets with a small active space (3zs, 4zs), we estimate the error of the basis
set on $\Psi^{\text{theo}}_\mathrm{0}$ to be below 1\,\%\  for the U state and below 6\,\%\ for the E' and B' states. Uncertainties from the overall size of active space, i.e. number of correlated electrons and virtual spinors, are estimated by comparison of 3zs and 3z calculations: errors are below 1\,\% for the U state, while errors for the E' and B' states are below 6\,\%. The uncertainty from the choice of intermediate and active spinors in the Fock space is estimated from the comparison of 3z and 3zl yielding uncertainties of 0.5\,\% for the U state and up to 9\,\% for the E' and B' states. Therefore, the total uncertainty from the active space and Fock space selection is 2\,\%\ for the U state and 12\,\%\ for the E' and B' states. 

By comparison to previous theoretical works \cite{tecmer:2018,oleynichenko:2023,zaitsevskii:2025} and experimental excitation energies, the effect of higher order coupled cluster amplitudes is estimated to be below 5\,\%\ for the U state and 12\,\%\ for the E' and B' states. The large uncertainty for the E' and B' states can be attributed to the pronounced configurational mixing of these states. Vibrational corrections are found to be negligibly small at the present level
of accuracy. The overall uncertainty on the computed molecular field shift factor is therefore estimated to be about 6\,\%\ for the U-X  and 17\,\% for the E'-X and B'-X transitions, resulting in the final values $\Psi_0^{\text{theo}}=-3.4(2)\,\mathrm{cm}^{-1}/\mathrm{fm}^2$ for U-X, $\Psi_0^{\text{theo}}=-2.5(4)\,\mathrm{cm}^{-1}/\mathrm{fm}^2$ for the E'-X and $\Psi_0^{\text{theo}}=-3.0(5)\,\mathrm{cm}^{-1}/\mathrm{fm}^2$ for the B'-X. This is in good agreement with the experimentally determined field shift $\Psi_0^{\text{exp.}} = -\SI{4.0\pm 1.3}{\cm^{-1} /fm^2}$ of the U-X transition. Ignoring the mass shift, the theory field shift together with the broad-band experimental isotopologue shift gives a $\langle r^2 \rangle^{229,232}$ of $\SI{0.308(18)}{fm^2}$ which agrees with the more precise ionic value of $\SI{0.316(9)}{fm^2}$ from Ref. \cite{Chakraborty2026}.

\section{Discussion and outlook}\label{sec:outlook}
We have demonstrated the successful implementation of the in-gas-jet laser ionization spectroscopy on cold molecules produced by laser ablation. The hypersonic nozzle was shown to efficiently cool the internal molecular degrees of freedom from a near room-temperature gas to a cryogenic gas jet, enabling rotationally resolved spectroscopy with just a few occupied ground state levels. This opens the door to precision studies of short-lived actinide molecules and establishes the suitability of the IGLIS technique for similar studies of shorter-lived molecules at on-line facilities.
\\
\\
In-gas-jet laser ionization spectroscopy was applied on thorium monoxide isotopologues, leading to the determination of the isotopologue shift in $^{229,230,232}$ThO. In parallel, laser-ablation targets were produced using thoriumtriflate and demonstrated to provide a reliable source of neutral ThO molecules. Ongoing developments aim at further increasing molecular production and improving target lifetime for extended experimental campaigns. While recoil sources may also provide access to these nuclei, efficient neutralization for subsequent resonance ionization spectroscopy remains one of the principal challenges. A promising route with recoil ions is to study molecules produced via, e.g., co-ablation, complementing previous demonstrations of laser spectroscopy on atomic recoil products in a gas jet \cite{Claessens2025}.
\\
\\
The measured rotational constants of $^{232}$ThO and $^{230}$ThO were found to agree with a pure change in inertial moment within the current experimental uncertainties. These uncertainties originate from the instabilities and statistics of the ablation process and importantly, the achievable line-width.
Apart from mass scaling, no isotope dependence could be resolved at the current level of precision, but measurements of Born-Oppenheimer breakdown terms can represent an important benchmark for future higher-accuracy studies.
\\
\\
A major outcome of this work is the combination of the measured molecular isotopologue shifts with high-accuracy isotope-shift data and calculations available for thorium (obtained in Th$^{3+}$ ions \cite{Zitzer2025, Chakraborty2026}) through a molecular King-plot analysis. This enabled the extraction of the molecular field-shift constant $\Psi_0$, which is up to 3 times larger than in ion transitions \cite{Chakraborty2026,Campbell2011,Zitzer2025,Safronova2018a}. The result indicates that molecular spectroscopy can provide significantly enhanced sensitivity to changes in the nuclear charge distribution and electron-nucleus interactions compared with atomic spectroscopy \cite{Athanasakiskaklamanakis2023}. Relativistic IH-FSCCSD calculations were used to determine the molecular constants $T_0$ and $B_0$, yielding values consistent within uncertainties with our experimental results. The calculated molecular field-shift constant \(\Psi^{\mathrm{theo}}\) for the U–X transition is also in good agreement with that extracted from the King-plot analysis, providing a stringent benchmark between  state-of-the-art relativistic many-body molecular and atomic calculations. 
\\
\\
This work lays the foundation for a new generation of measurements on cold radioactive molecules at the IGLIS laboratory. Ongoing narrow-band laser spectroscopy of $^{229}$ThO is expected to improve by more than an order of magnitude the precision of the isotope-shift measurements, allowing more accurate determinations of both the field- and mass-shift constants. The reported results represent an important first step towards measuring the hyperfine structure of $^{229}$ThO that will provide direct additional access to nuclear magnetic dipole and electric quadrupole moments of this unique (nuclear-clock) nucleus. Besides a key benchmark for ab-initio molecular calculations against atomic hyperfine field and electric field gradient calculations, such a study probes the local electronic structure in the vicinity of the $^{229}$Th nucleus. This allows us to validate calculations of symmetry-violating matrix elements which are not otherwise experimentally accessible \cite{Safronova2018}
\\
\\
More generally, the compatibility of the IGLIS technique with on-line operation offers a clear pathway towards routine production and spectroscopy of radioactive molecules at present and upcoming heavy-ion accelerator facilities such as JetRIS at GSI (Darmstadt, Germany)  \cite{Lantis2024}, S$^3$-LEB at GANIL (Caen, France) \cite{AJAYAKUMAR2023102} and SPECTRE at FRIB (East Lansing, Michigan, USA). Based on this work, a dedicated off-line IGLIS setup is under development at Harvard specifically for thorium-containing molecules. The introduction of reagent gases, such as O$_2$ or CF$_4$ \cite{Dewberry2007}, or laser co-ablation of reagent materials \cite{Conn2026}, should enable efficient molecular formation directly inside the gas cell. To this end, to demonstrate the formation of the heaviest actinide diatomic molecules (AnX),  initial studies will investigate these processes under off-line conditions using lighter homologues of the molecules of interest, such as Yb (No) and Lu (Lr), to produce possible compounds such as NoF and LrO. Spectroscopy of nobelium molecules would complement current atomic studies as well as could benefit from the possible enhanced sensitivity of molecular transitions to nuclear-size effects demonstrated in this work for ThO, while the formation of volatile molecular compounds may ultimately provide the only practical route to spectroscopy of lawrencium; repeated attempts to identify atomic transitions in $^{255}$Lr via filament desorption have so far remained unsuccessful \cite{Rickert2026}.
\\
\\
All in all, the results presented here establish in-gas-jet laser ionization and spectroscopy as a powerful tool for the production and investigation of cold actinide molecules. By combining efficient cooling, overall efficiency, high spectral resolution, molecular production, low spectral density and compatibility with on-line radioactive-beam facilities, the IGLIS technique opens a broad experimental program spanning nuclear structure, atomic and molecular theory, along with necessary preparatory precision tests of fundamental interactions in the heaviest systems.

\section{Acknowledgments}
KG gratefully acknowledges funding by the Fonds der
Chemischen Industrie through a Liebig fellowship, computing time provided on the HIMSter III supercomputer at the Helmholtz Institute Mainz (HIM) and the Johannes Gutenberg University Mainz, and the Center for Information Technology of the University of Groningen for their support and for providing access to the Hábrók high performance computing cluster.
\\
\\
This work has received funding from European Union’s Horizon
2020 research and innovation programme under Grant
Agreement No. 861198-LISA-H2020-MSCA-ITN-2019, the
Research Foundation Flanders (FWO, Belgium) with grants G078624N and G0814522N, as well as the International Research Infrastructure (IRI) Project No. I001323N, BOF KU
Leuven (C14/22/104), and the FWO and F.R.S.-FNRS under
the Excellence of Science (EOS) program project nr. 40007501.

\bibliography{sources_article_molecules}

\end{document}